\documentclass[11.pt]{article}

\usepackage{latexsym}
\usepackage{epsfig}
\usepackage{color}
\usepackage{caption}
\def\R{ {\rm R \kern -.31cm I \kern .15cm}}
\def\C{ {\rm C \kern -.15cm \vrule width.5pt \kern .12cm}}
\def\Z{ {\rm Z \kern -.27cm \angle \kern .02cm}}
\def\N{ {\rm N \kern -.26cm \vrule width.4pt \kern .10cm}}
\def\1{{\rm 1\mskip-4.5mu l} }
\def\lsim{\raise0.3ex\hbox{$<$\kern-0.75em\raise-1.1ex\hbox{$\sim$}}}
\def\gsim{\raise0.3ex\hbox{$>$\kern-0.75em\raise-1.1ex\hbox{$\sim$}}}

\def\beq{\begin{equation}}   \def\eeq{\end{equation}}
\def\bea{\begin{eqnarray}}  \def\eea{\end{eqnarray}}

\DeclareGraphicsExtensions{.eps}

\def\lsim{\raise0.3ex\hbox{$<$\kern-0.75em\raise-1.1ex\hbox{$\sim$}}}
\def\gsim{\raise0.3ex\hbox{$>$\kern-0.75em\raise-1.1ex\hbox{$\sim$}}}

\newcommand{\red}[1]{\textcolor[rgb]{0,0,0}{#1}}

\begin{document}

\title{\bf Open charm production in high multiplicity proton-proton events at the LHC}

\vskip 8. truemm
\author{\bf E. G. Ferreiro and C. Pajares}
\vskip 5. truemm

\date{}
\maketitle

\begin{center}
\small{
  Departamento de F{\'\i}sica de Part{\'\i}culas and IGFAE, Universidad de
  Santiago de Compostela, \\
  15782 Santiago de Compostela, Spain}
\end{center}
\vskip 5. truemm

\begin{abstract}
We present the dependence of $D$ production on the charged particle multiplicity in proton-proton collisions at LHC energies.
We show that, in a framework of source coherence, 
the open charm production exhibits a growth with the multiplicity
which is stronger than linear in the high density domain.
This departure from linearity was previously observed in the $J/\psi$ \red{inclusive} data from proton-proton collisions at 7 TeV 
and was successfully described in our approach.
Our assumption, the existence of coherence effects present in proton-proton collisions at high energy,
applies for high multiplicity proton-proton collisions in the central rapidity region and
should affect any hard observable.
\end{abstract}
\vskip 3 truecm

\hspace*{\parindent}
\pagestyle{myheadings}
\section{Introduction}
The available data recoiled at 
RHIC \cite{Back:2004je,Adcox:2004mh,Adams:2005dq,Arsene:2004fa} and LHC \cite{Schukraft:2011cz,Steinberg:2011dj,Wyslouch:2011zz} 
on heavy-ion collisions have already shown several features
which indicate the formation of a high-density partonic medium.
In which concerns $pp$ collisions at LHC energies, the energy densities achieved in the high multiplicity events 
are similar to the ones reached in CuCu central collisions or AuAu peripheral collisions at $\sqrt{s_{NN}}=200$ GeV.
It is then mandatory to look for observables which could reflect the formation of a high density medium, similar to the one already observed in heavy-ion collisions.

\vskip 0.25cm
One of those signals, already reported by the ALICE collaboration \cite{Abelev:2012rz,PorteboeufHoussais:2012gn},
was the rise of  \red{central rapidity} $J/\psi$ \red{inclusive} production in the highest multiplicity events obtained in $pp$ collisions at 7 TeV, that was successfully reproduced in the source interaction framework \cite{Ferreiro:2012fb}.

Here we address to the production of open charm mesons. We will restrict ourselves to the study of the lightest ones, $D^0$, $D^+$ and $D^-$, and we will 
comment on the other states.

\section{The model}

Let us recall the essential ingredients of our model.
Our main assumption consists on considering high-energy hadronic collisions as driven by the exchanged of 
colour sources --strings-- between the projectile and the target. Those sources have finite spatial extension and thus they can interact,
reducing their effective number.

\vskip 0.25cm
Note that one can distinguish between soft and hard sources, depending on their transverse mass, i.e. their quark composition and their transverse momentum. 
Note also that their transverse size is determined by this transverse mass, since $r_T \ \propto \ 1/m_T$.
The softness of the source maximizes its possibility of interaction, since its transverse size will be larger.

\vskip 0.25cm
The main consequence is that the bulk properties --as the total multiplicities-- that are driven by the soft sources are going to be affected by their initial interactions, in particular reducing the total number of produced particles in similar way as shadowing or saturation do. On the other hand, the hard events are less affected by source interaction and can be taken as proportional to the number of collisions. 

\vskip 0.25cm
Specifically, in our model each parton-parton collision is reflected as the number of initially produced sources $N_s$.
The hard events in $pp$ collisions are mainly controlled by this number, i.e. the number of collisions.
On the contrary, the multiplicity distribution $dN/d\eta$ --mainly soft-- 
is not proportional to the number of collisions, but mostly to the number of participants. This reduction can be considered as a consequence of shadowing in the case of $pA$ or $AA$ collisions \cite{Ferreiro:2008wc}, parton saturation \cite{Kharzeev:2008nw} or, as it is the case here, string interactions \cite{Armesto:1996kt}. 
As a consequence,
the charged particle multiplicities can suffer a reduction due to the interaction among the sources
and behave roughly as $\sqrt{N_s}$. 
The above assumptions are similar to the fact that the shadowing effects 
decrease with the hardness of the probe.

\vskip 0.25cm
In particular, in our approach the multiplicity distribution is given by
\beq
\frac{dN}{d\eta} = F(\rho) N_s \mu_1
\label{eq1}
\eeq
where $\mu_1$ corresponds to the multiplicity of a single source in the rapidity range of interest, $N_s$ is the number of produced sources and $F(\rho)$ corresponds to the 
damping 
factor induced by the source interaction, 
\beq
F(\rho)=\sqrt{\frac{1-e^{-\rho}}{\rho}} \, .
\label{eq2}
\eeq
$\rho$ corresponds to the source density,
$\rho=\frac{N_s \sigma_0}{\sigma}$, being $\sigma_0$ the transverse size of one source and
$\sigma$ the transverse area of the collision. 

\vskip 0.25cm
In our previous paper \cite{Ferreiro:2012fb} we have assumed
the proportionality between the number of produced $J/\psi$ and the number of
collisions,
\beq
\frac{n_{J/\psi}}{<n_{J/\psi}>}=\frac{N_s}{<N_s>} \, ,
\label{eq3}
\eeq
obtaining 
the relation between the charged particle multiplicity and the number of produced $J/\psi$, 
\beq
\frac{\frac{dN}{d\eta}}{<\frac{dN}{d\eta}>} = \left ( \frac{n_{J/\psi}}{<n_{J/\psi}>} \right )^{1/2} 
\left [ \frac{1-e^{-\frac{n_{J/\psi}}{<n_{J/\psi}>}<\rho>}}{1-e^{-<\rho>}} \right ]^{1/2} \, {\rm where}\, 
<\rho>=<N_s> \frac{\sigma_0}{\sigma}\, .
\label{eq4}
\eeq
The above equation leads to the following behaviour for the $J/\psi$ production:
At low multiplicities, where the number of sources $<N_s>$ is small, one obtains the linear 
dependence 
\beq
\frac{n_{J/\psi}}{<n_{J/\psi}>}=\frac{\frac{dN}{d\eta}}{<\frac{dN}{d\eta}>} \, ,
\label{eq5}
\eeq
while, at high multiplicities, 
\beq
\frac{n_{J/\psi}}{<n_{J/\psi}>}=<\rho> \left ( \frac{\frac{dN}{d\eta}}{<\frac{dN}{d\eta}>} \right )^2 \, ,
\label{eq6}
\eeq
where the linear dependence changes to an squared dependence when high multiplicity events are at play.
This can be parametrised by the following expression \cite{Ferreiro:2012fb}:
\beq
\frac{n_{J/\psi}}{<n_{J/\psi}>}= (1-<\rho>) \left ( \frac{\frac{dN}{d\eta}}{<\frac{dN}{d\eta}>} \right ) + <\rho> \left ( \frac{\frac{dN}{d\eta}}{<\frac{dN}{d\eta}>} \right )^2 \, ,
\label{eq7}
\eeq
which agrees extremely well the available experimental data \cite{Abelev:2012rz,PorteboeufHoussais:2012gn}.

\vskip 0.25cm
Let us now face the problem of open charm production. We have considered that, in first approximation, $D$ production is driven both by light and hard sources, due to their mixed composition. Both contributions are to be taken at $50 \%$. The soft contribution will follow the linear behaviour with the multiplicity, while the hard one will behave as the $J/\psi$ does, according to eq. (\ref{eq7}),
\beq
\frac{n_{D}}{<n_{D}>}= 0.5 \left ( \frac{\frac{dN}{d\eta}}{<\frac{dN}{d\eta}>} \right ) +
0.5 \left [ (1-<\rho>) \left ( \frac{\frac{dN}{d\eta}}{<\frac{dN}{d\eta}>} \right ) + <\rho> \left ( \frac{\frac{dN}{d\eta}}{<\frac{dN}{d\eta}>} \right )^2 \right ] \, .
\label{eq8}
\eeq
%
%
\vskip -0.25cm
\begin{table}[htb!]
\begin{center}\setlength{\arrayrulewidth}{1pt}
\caption{Relative charged-particle pseudorapidity densities, mean source densities, and relative $J/\psi$ and $D$ yields according to eqs. (7,8).}
\label{tab1}
\begin{tabular}{ccccccc}
\hline\hline
 $\frac{\frac{dN}{d\eta}}{<\frac{dN}{d\eta}>}$ & $<\rho>$ & $\frac{n_{J/\psi}}{<n_{J/\psi}>}$ & $\frac{n_{D}}{<n_{D}>}$ \\
\hline
1 & 0.09 & 1.00 & 1.00 \\
2 & 0.16 & 2.32 & 2.16 \\
3 & 0.23 & 4.37 & 3.69 \\
4 & 0.29 & 7.46 & 5.73 \\
5 & 0.34 & 11.88 & 8.44 \\
6 & 0.40 & 17.88 & 11.94 \\
7 & 0.46 & 26.15 & 16.58 \\
8 & 0.51 & 36.73 & 22.36 \\
9 & 0.58 & 50.69 & 29.84 \\
\hline\hline
\end{tabular}
\end{center}%
\vspace*{-0.5cm}
\end{table}

\vskip 0.25cm
In order to compare with the available $pp$ experimental data, 
we need to compute the dependence of the source densities with the multiplicities. 
In Table 1, we show our values for the 
available experimental beams together with our results for $J/\psi$ and $D$ production according to eqs. (\ref{eq7},\ref{eq8}).
Note that the $D$ meson results shown in the above table corresponds to the lightest ones, i.e. $D^0$, $D^+$, $D^-$, and are to be taken as $p_T$ integrated, i.e. $p_T > 0$ GeV. 

\vskip 0.25cm
The experimental results on open charm production are presented for different species and different $p_T$ cuts.
In order to compare with experimental data \cite{Zaida}, we have taken the lowest $p_T$ beam, i.e. $2<p_T<4$ GeV for $D^0$ and $D^+$ production. In Fig. 1, we show our results compared to experimental ALICE data \cite{Zaida} for $D$ meson production in $pp$ 
collisions at 7 TeV in the central rapidity range.

\vskip 0.35cm
\begin{figure*}[htb!]
\begin{center}
\includegraphics[width=0.75\textwidth]{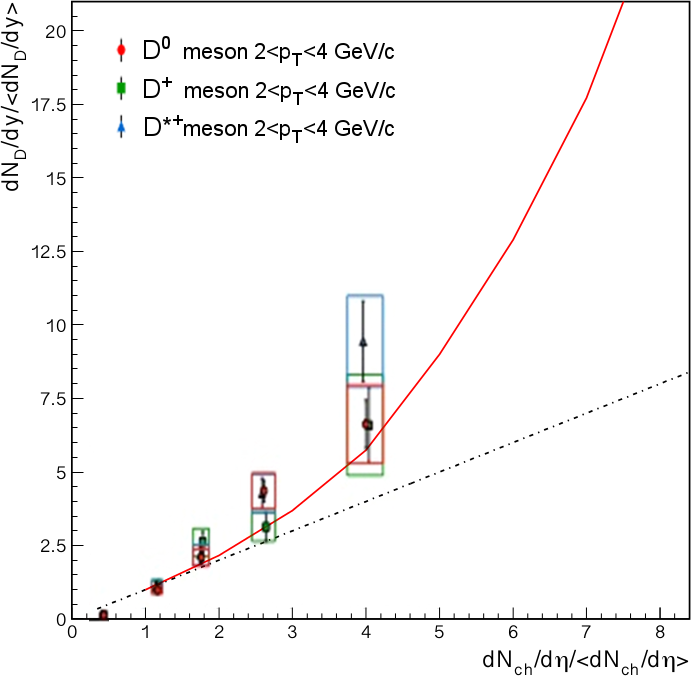}
\end{center}
\caption{
Our results on $D$ meson production (red line) for $pp$ collisions in the central rapidity range, 
together with the experimental data 
from the ALICE Collaboration \cite{Zaida}. 
The linear behaviour
(black line) is also plotted.}
\label{fig1}
\end{figure*}
We observe a good agreement between our result and the experimental data for $D^0$ and $D^+$ production.
Note that, in which concerns $D^*$ production, 
its departure from linearity is to be more important, due to its higher mass.
Moreover, we can advance that the departure from linearity should also
be more important for the high $p_T$ beams when compare to the low $p_T$ beams.

\vskip 0.25cm
In Fig. 2, our results on $D$ and $J/\psi$ are plotted and compared to experimental data.

\vskip 0.35cm
\begin{figure*}[htb!]
\begin{center}
\includegraphics[width=0.75\textwidth]{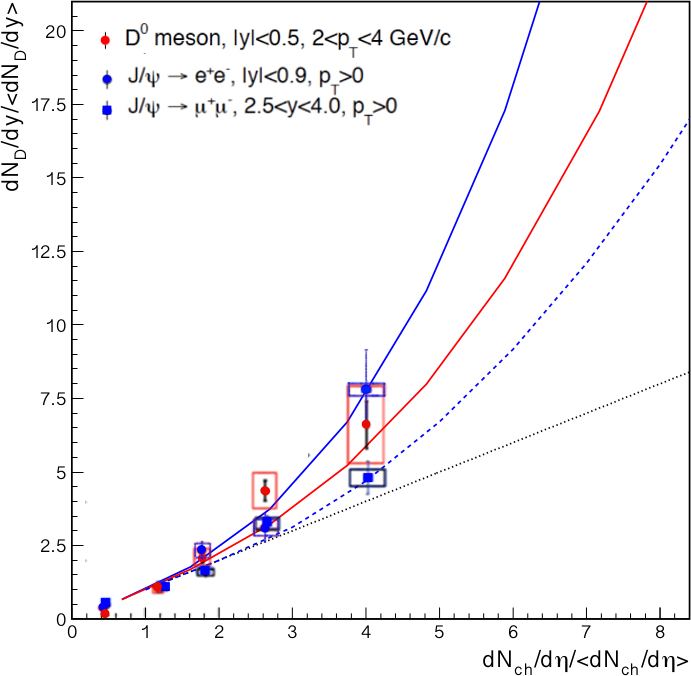}
\end{center}
\caption{
Our results on $D$ meson production in the central rapidity range (red line) and on $J/\psi$ production in the central (continuous blue line) and in the forward (dashed blue line) rapidity range for $pp$ collisions at 7 TeV,
together with the experimental data
from the ALICE Collaboration \cite{Zaida}.
The linear behaviour
(black line) is also plotted.}
\label{fig2}
\end{figure*}
Note that the $J/\psi$ production obeys a quadratic dependence with the multiplicity in the central rapidity region, while it is closer to linear dependence in the forward rapidity one
\cite{Ferreiro:2012fb}. 


\section{Conclusions}
In conclusion, we have reproduced here the rise of 
$D$ and $J/\psi$ production for the highest multiplicity events at central rapidity observed by the ALICE collaboration in $pp$ collisions. 
This increase may be a consequence of the formation of a 
high density medium in $pp$ collisions at LHC energies. 

\vskip 0.25cm
\red{
At these high densities, the coherence among the sources can lead to a reduction of their effective number --initially proportional to the number of collisions. This reduction would mainly affect the soft observables, as the total multiplicity, while the hard production would remain unaltered.}
In this case, the linear dependence of $D$ and $J/\psi$ production on the charged particle multiplicity obtained for low multiplicities 
--where the parton densities are smaller--, changes to an squared dependence 
when high multiplicity events are at play.
%
\red{Moreover, this departure from linearity should increase with the hardness of the studied observable. This means that, for the same $p_T$ and rapidity range, the $J/\psi$ production will be higher than the $D^0$ production for the high multiplicity events. For the same particle specie there should be also an ordering in $p_T$, i.e. the highest $p_T$ particles should show a most important departure of linearity.}  

\vskip 0.25cm
Note that any additional $J/\psi$ suppression --due to the possibility of string or source percolation--, which could lead to a corresponding increase of the $D$ production, cannot be at play at these energies and multiplicities, since the density of strings is well below the threshold for percolation, $\eta > 1$.
\red{
This threshold can ben nevertheless achieved in $pp$ collisions at higher energies, i.e. 14 TeV, for the high multiplicity events, $\frac{\frac{dN}{d\eta}}{<\frac{dN}{d\eta}>} > 8$ and also in $p$Pb collisions at 5.02 TeV.
In the latest case, the application of our model is not completely straightforward, since other cold nuclear matter effects can be at play here.}

\vskip 0.35cm
{\it Acknowledgements.---}
We are grateful to Zaida Conesa del Valle and Sarah Porteboeuf-Houssais  for useful
discussions.
This work is supported by
Ministerio de Economia y Competitividad 
of Spain 
and FEDER.

\end{document}